\documentclass[11pt]{article}

\usepackage[utf8]{inputenc}
\usepackage[T1]{fontenc}
\usepackage{lmodern}
\usepackage[english]{babel}
\usepackage[a4paper,margin=2.5cm]{geometry}
\usepackage{graphicx}
\usepackage{booktabs}
\usepackage{array}
\usepackage{amsmath,amssymb}
\usepackage{enumitem}
\usepackage{caption}
\usepackage{longtable}
\usepackage{textcomp}
\usepackage[hidelinks]{hyperref}
\usepackage{xurl}

\newcolumntype{P}[1]{>{\raggedright\arraybackslash}p{#1}}

\title{\bfseries Enhancing Communication in Speech Therapy:\\
Exploring the Cognitive Synergy Between Gesture and Speech}

\author{%
  Paul-Peter Arslan \quad Xiao Xiao\\[2pt]
  \normalsize De Vinci Research Center, De Vinci Higher Education, Paris, France\\
  \normalsize \texttt{paulpeterarslan@gmail.com}
}

\date{2026}

\begin{document}

\maketitle

\begin{abstract}
\noindent
This paper examines the adaptation of a rhythm-based interface, originally designed for manual
dexterity rehabilitation, for use in speech therapy. The interface allows users to control
synthesized vocal phrases through finger tapping, leveraging the cognitive link between gesture
and speech. Through interviews with four therapists and pilot tests with one speech therapist and
eight children with speech impairments (autism, Down syndrome, verbal apraxia, dyslexia), we found
that the interface improves motivation and therapeutic outcomes by facilitating more interactions
between verbally challenged patients and the therapist. Our findings suggest that rhythmic gestures
can enhance verbal communication, offering potential for broader therapeutic and educational
applications.
\end{abstract}

\emph{Auditory feedback}; \emph{User studies}.


\vspace{1em}

\begin{figure}[htbp]
  \centering
  \includegraphics[width=\textwidth]{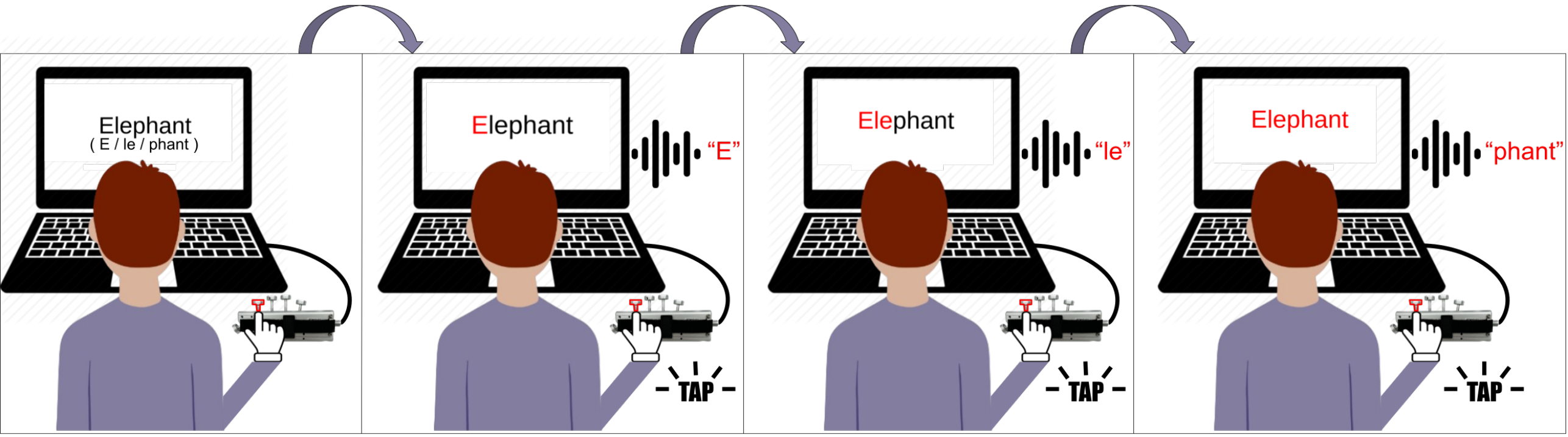}
  \caption{The user interacts with the interface by tapping a button on the Dextrain Manipulandum,
  aligning finger movements with vocal articulation. The word ``Elephant'' is displayed on the
  screen and is progressively highlighted in red syllable by syllable each time the button is
  pressed (``E,'' ``le,'' ``phant'').}
  \label{fig:teaser}
\end{figure}

\clearpage

\section{Introduction}

Rhythm is vital in a variety of rehabilitation therapies, particularly for enhancing movement
coordination and synchronization~\cite{thaut2010rhythmic}. In physical therapy, rhythmic exercises
are crucial for improving motor skills and body awareness, especially for patients recovering from
brain injuries or strokes~\cite{ghai2018effects}. Similarly, rhythm plays a key role in speech
therapy, where it aids in improving fluency, intonation, and articulation~\cite{zumbansen2014melodic}.

Our research investigates the cognitive link between hand gestures, rhythm, and speech through an
interactive system that allows users to control the rhythm of a synthesized voice via finger
tapping~\cite{luo2023studying}. Originally developed for stroke patients to improve manual
dexterity~\cite{teremetz2015novel,teremetz2023efficacy}, the interface demonstrates how integrating
hand gestures with speech rhythm can enhance both performance and motivation.

We aim to adapt this system for broader applications. As a preliminary step, we conducted
semi-structured interviews with therapists, including two physical therapists and two speech
therapists. We found speech therapy particularly promising, given that hand gestures often accompany
verbal communication~\cite{andric2012gesture}. We then explored how rhythmic gestures might aid in
rehabilitating speech disorders by gathering user feedback on the interface with eight children who
have various speech disorders~\cite{braun2006using,braun2016using}.

Our results indicate that the interface effectively links motor skills with speech, improving therapy
outcomes and patient motivation. Therapists noted that using gestures in combination with rhythms
helped children better memorize and pronounce words. The research also highlighted the need for a
customizable interface, leading to refinements that were personally adapted for each patient.

This interface, initially designed for motor rehabilitation, was shown to be effective in speech
therapy, demonstrating that the connection between gestures and speech is beneficial. The technology
shows potential for broader therapeutic applications, suggesting that rhythmic integration in therapy
could significantly improve patient outcomes in both motor and speech
rehabilitation~\cite{braunjanzen2022rhythm}. This research opens the door for further exploration and
the development of adaptable therapeutic tools~\cite{mortenson2015development}.

\section{Related Work}

\subsection{The Role of Rhythm in Rehabilitation}

Rhythm is central in motor rehabilitation, improving timing, coordination, and motor learning in
neurological patients of strokes or Parkinson's
syndrome~\cite{ashoori2015effects,nieuwboer2007cueing,nombela2013into,schaefer2014auditory,thaut1996rhythmic}.
By engaging brain regions such as the premotor cortex and supplementary motor area, rhythmic
interventions enhance cortical plasticity and motor synchronization, leading to better motor
recovery~\cite{bengtsson2009listening,johansenberg2002attention,rosenkranz2007motorcortical}.
Repetitive rhythmic patterns accelerate motor learning by incorporating timing into neural
circuits~\cite{bassett2015learning,dayan2011neuroplasticity,grafton1996localization,hikosaka2002central}.
In speech therapy, rhythm aids fluency and coordination~\cite{peelle2012neural}, with
metronome-based interventions improving speech production, particularly in children with verbal
apraxia~\cite{mainka2014rhythmic}. Rhythm provides a temporal framework that helps patients better
organize speech patterns, improving articulation and
fluency~\cite{kotz2010cortical,patel2014evolutionary,peelle2012neural}.

Rhythmic tools like Rhythmic Auditory Stimulation (RAS) and the Dextrain
Manipulandum~\cite{teremetz2015novel} are effective in motor
recovery~\cite{braunjanzen2022rhythm,gonzalez2021effects}. RAS is a therapeutic technique that uses
rhythmic auditory cues, such as metronome beats, to improve movement timing and coordination, often
used in gait training for neurological patients. The Dextrain Manipulandum is a hardware device
designed to measure the pressure and movement of each finger individually, facilitating exercises
that improve manual dexterity, finger strength, and temporal precision through fine motor control
tasks~\cite{teremetz2023efficacy}. Metronome-based systems, which provide rhythmic cues, are also
useful in treating speech disorders like stuttering and verbal apraxia, helping organize speech
sequences and improve fluency~\cite{mainka2014rhythmic}.

\subsection{How Gestures and Speech are Connected}

Research shows a strong cognitive link between gestures and
speech~\cite{kendon2004gesture,mcneill2011gestures,zhao2023tms}, where gestures are an integral part
of language production~\cite{hostetter2019gesture,kelly2015processing,kita2017how}. The ``Growth
Point Theory'' posits that gestures and speech arise from a shared cognitive framework, with gestures
representing visuospatial aspects of linguistic content~\cite{mcneill2005gesture}. Neuropsychological
evidence supports this connection, indicating overlapping neural circuits for speech and gestures,
treating them as part of the same communicative act~\cite{dick2012gesture,skipper2007speech}.
Gesture-based therapies have been successfully used in treating neurogenic communication disorders
like aphasia~\cite{sekine2013relationship}.

\subsection{New Therapies Combining Rhythm and Gestures}

Combining rhythm with gesture-based therapies has significant potential for speech rehabilitation,
particularly in patients with speech disorders. Studies show that rhythmic synchronization with
gestures enhances word recall and pronunciation in these
populations~\cite{fujii2014role,nirme2024early}. This approach leverages the brain's natural
multimodal integration, where motor actions (gestures) and linguistic elements (speech) are processed
together, enhancing learning and retention~\cite{fujii2014role,pulvermuller2005brain}. Gesture-based
interfaces, which enhance motivation and engagement through interactive therapy, have shown promise
for patients with autism and Down syndrome~\cite{delrio2019hand,keaybright2011cocreating}. Combined
with rhythmic repetition, they also improve articulation and fluency by supporting cognitive and
motor functions~\cite{fujii2014role}. These results highlight the importance of integrating both
rhythm and gestures in therapeutic interventions to effectively support speech rehabilitation.

While current research shows cognitive and neurological benefits of combining rhythm and gestures in
rehabilitation, there are significant opportunities for future development. Gesture-rhythm tools
could be expanded in educational and therapeutic contexts, with future research focusing on more
interactive, adaptive gesture-speech interfaces across a wider range of conditions.

\section{Development \& Description of the Interface}

\subsection{Interface Design}

\begin{figure}[htbp]
  \centering
  \includegraphics[width=0.72\textwidth]{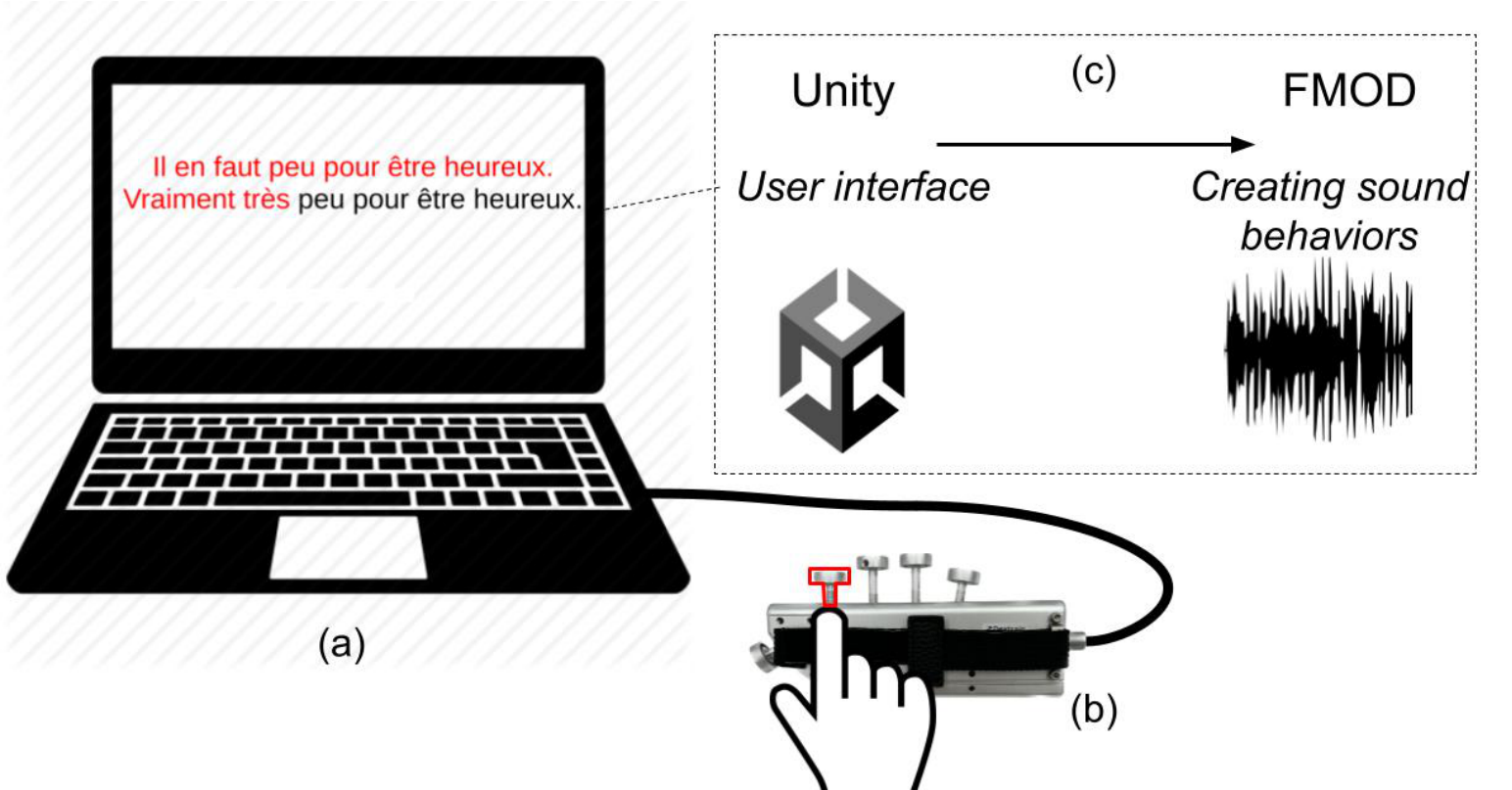}
  \caption{Illustration of the interface system. (a) The user interface displays black words on the
  screen, with each syllable turning red as the user presses a button, accompanied by an audible
  output. (b) The Dextrain Manipulandum~\cite{teremetz2015novel} is used for tapping, where the
  participant presses a piston with their index finger, recording precise movements and timing.
  (c) The software, developed in Unity with FMOD, controls the synthesis of vocal and musical sounds.}
  \label{fig:system}
\end{figure}

Our system is built as an application for the Dextrain Manipulandum
(DM)~\cite{teremetz2015novel}, a device that accurately measures the force exerted by each finger,
originally designed for rehabilitating manual dexterity in stroke patients. In this study, the DM is
adapted to assist users in improving both their speech and fine motor skills through rhythmic
repetition exercises (see Figure~\ref{fig:system}). The platform allows users to replicate words and
phrases they hear, with a focus on rhythmic accuracy and articulation. The system combines the DM
with a Unity program and the FMOD audio interface~\cite{robinson2019game} to generate synthesized
speech segments, ensuring precise timing and optimal sound clarity.

\subsection{Specific Features}

\textbf{User Interface and Hardware.} Users see the words they need to reproduce displayed on the
screen in black. With each button pressed by their index finger, a syllable of the word is audibly
generated, and the corresponding syllable on the screen turns red. This allows users to visually and
audibly track their progress, reinforcing the association between movement, sound, and speech.

\textbf{Sound Synthesis and Control.} The sound is generated using pre-recorded audio files, with
each word segmented into syllables with precise timestamps. When a user presses the button, the
corresponding syllable is played, helping to maintain a consistent rhythm and practice correct
pronunciation. This method ensures that users can focus on synchronizing their movements with sound
production, thereby improving their rhythmic accuracy and articulation.

\begin{figure}[htbp]
  \centering
  \includegraphics[width=0.8\textwidth]{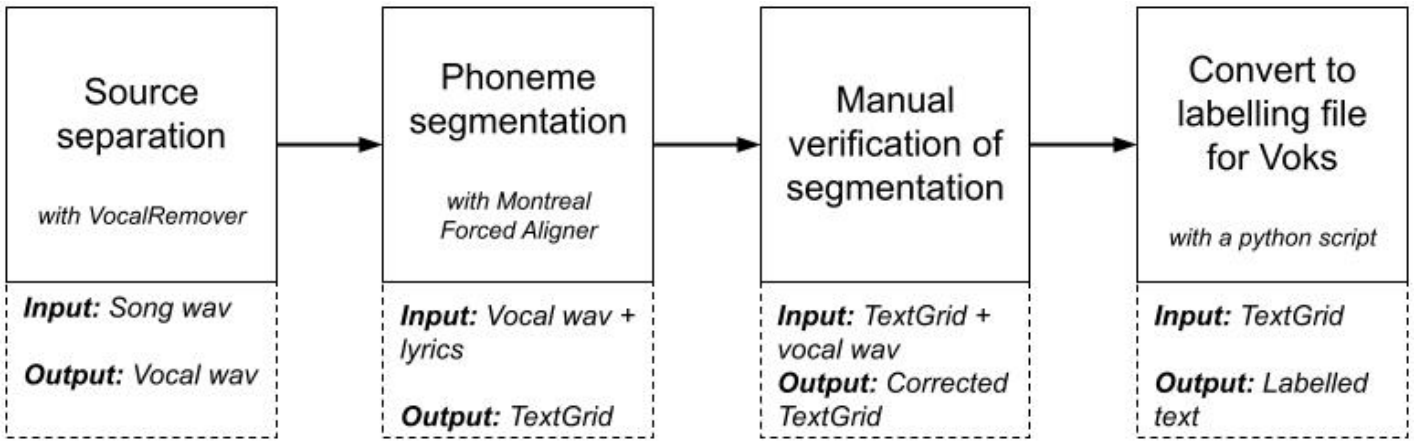}
  \caption{Data preparation pipeline.}
  \label{fig:pipeline}
\end{figure}

\begin{figure}[htbp]
  \centering
  \includegraphics[width=0.85\textwidth]{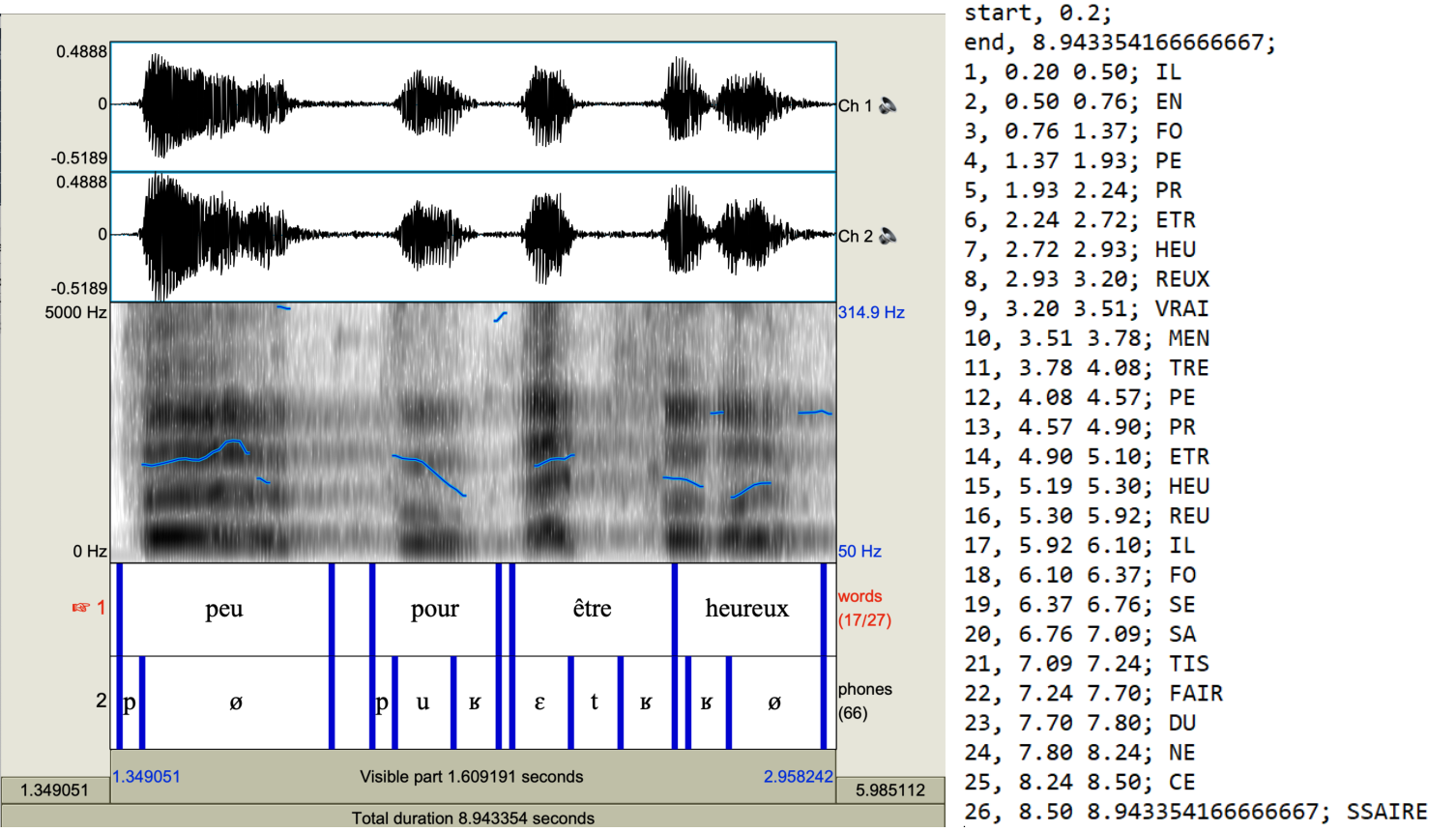}
  \caption{(Left) Syllable segmentation generated by Montr\'eal Forced
  Aligner~\cite{mcauliffe2017montreal} opened in Praat~\cite{boersma2022praat} for manual correction.
  (Right) Labeling file generated by the Python script. Letters to be highlighted in the Unity
  interface are manually added.}
  \label{fig:praat}
\end{figure}

\textbf{Data Preparation.} To prepare the speech content for use, the moments of transition between
different phonemes need to be labeled accurately. To achieve precise timing for the speech exercises,
we developed a semi-automatic pipeline for processing vocal tracks in \texttt{.wav} format (see
Figure~\ref{fig:pipeline}). First, we separate the vocals from the instrumental background using
Vocal Remover~\cite{vocalremover}, an online AI-based tool. Next, we segment the phonemes using
Montr\'eal Forced Aligner~\cite{mcauliffe2017montreal}, which generates a Praat TextGrid file
containing the start and end timestamps of each phoneme~\cite{boersma2022praat}. We manually verify
and correct these timestamps in Praat to ensure accurate synchronization (see
Figure~\ref{fig:praat}). Finally, a Python script converts the TextGrid to the required format for
our interface. The correctly labeled speech segments are then integrated into the exercises.

\section{Methodology}

\subsection{Participants}

This study involved four therapists: two speech therapists (T1, T2) and two physical therapists
(T3, T4) (see Table~\ref{tab:therapists}). All four therapists took part in interviews, while the
testing sessions with children were conducted exclusively with T1, a speech therapist specializing in
pediatric speech therapy. The children (P1--P8), aged between six and twelve, were already receiving
therapy with T1 and had various speech disorders, including autism, Down syndrome, verbal apraxia,
and dyslexia (see Table~\ref{tab:children}).

\begin{table}[htbp]
  \centering
  \caption{Demographic data of therapist participants.}
  \label{tab:therapists}
  \small
  \setlength{\tabcolsep}{5pt}
  \begin{tabular}{lccccl}
    \toprule
    Therapist (T) & Gender & Age & Specialty & Years of Exp. & Focus Area \\
    \midrule
    T1 & Female & 35 & Speech Therapist & 11 & Pediatric Speech Pathologies \\
    T2 & Female & 27 & Speech Therapist & 4  & Post-Stroke Speech Rehabilitation \\
    T3 & Female & 39 & Physiotherapist  & 14 & Motor Rehabilitation \\
    T4 & Male   & 60 & Physiotherapist  & 35 & Parkinson's Disease Rehabilitation \\
    \bottomrule
  \end{tabular}
\end{table}

\begin{table}[htbp]
  \centering
  \caption{Demographic data of child participants.}
  \label{tab:children}
  \small
  \begin{tabular}{lccl}
    \toprule
    Participant (P) & Gender & Age & Pathology \\
    \midrule
    P1 & Female & 6  & Autism \\
    P2 & Male   & 7  & Down Syndrome \\
    P3 & Female & 8  & Verbal Apraxia \\
    P4 & Male   & 9  & Dyslexia \\
    P5 & Female & 10 & Autism \\
    P6 & Male   & 11 & Verbal Apraxia \\
    P7 & Female & 12 & Dyslexia \\
    P8 & Male   & 6  & Down Syndrome \\
    \bottomrule
  \end{tabular}
\end{table}

\subsection{Procedures}

The study was structured into three phases: (1) interviews and a demonstration of the rhythmic
interface, (2) implementing adaptations based on interview feedback, and (3) testing sessions with
the speech therapist and her eight child patients.

\subsubsection*{4.2.1 Interviews and Demonstration}

The interviews with the therapists were divided into three main sections. First, we inquired about
their therapeutic practices, specifically how they incorporate speech, rhythm, and gestures into
their sessions. Next, we demonstrated the interface to familiarize them with its functionality.
Finally, we collected their input on how the interface could be adjusted to better suit their needs.
They also provided suggestions on incorporating new content and establishing criteria for evaluating
the tool's effectiveness.

Interview questions were predefined as open questions to ensure the collection of a wide range of
responses~\cite{ruud2023patient}, and they were asked in a consistent manner across all participants
to maintain comparability. The questionnaire used in the interviews can be found in
Appendix~\ref{app:protocol}. All responses were recorded and transcribed verbatim for later analysis.

The purpose was to explore the connection between gestures and speech in both speech and physical
therapy and to determine how the tool could support both disciplines. This phase also helped us
define how we would assess the interface's impact in practice.

\subsubsection*{4.2.2 Implementation of Adaptations}

Based on the feedback from the interviews, we made specific changes to the interface to ensure it
aligned with the therapists' needs. These adaptations focused both on improving the tool's usability
in real-world therapy sessions and on adding exercises tailored to their specific therapeutic
practices.

\subsubsection*{4.2.3 Testing Sessions}

The testing sessions were conducted over two days, with T1 supervising eight children (P1--P8) during
30-minute sessions. The children interacted with the adapted interface, and we observed their
engagement and progress in speech exercises. T1 provided additional feedback on how the tool
functioned in practice and noted any challenges or improvements seen in the children's performance.

\subsection{Data Collection and Analysis}

Data were collected through several methods to evaluate the interface. All therapist interviews were
audio-recorded, transcribed, and analyzed using thematic
analysis~\cite{braun2006using,braun2016using}. This process helped identify patterns related to the
integration of the tool into their practices, as well as commonalities between the therapists'
approaches, particularly regarding the connection between gestures and speech. This analysis also
highlighted areas where the interface could be adapted and improved.

During the testing sessions, observations focused on the children's interactions with the interface,
including their engagement, motivation, and progress in speech exercises. The supervising speech
therapist (T1) provided feedback on each child's response to the tool. These observations and
evaluations were analyzed to understand the interface's impact on the children's performance.

All data were anonymized, and consent was obtained from both therapists and the children's parents.
The coding and analysis were conducted by a single researcher and ethical guidelines were followed
throughout the study.

\section{Findings}

\subsection{Specific Challenges in Sessions}

Rehabilitation sessions in both speech and physical therapy are complex and require constant
adaptation. Therapists must adjust their approaches to meet the varying needs and behaviors of
patients, which can fluctuate even within a single session. In speech therapy, maintaining a child's
engagement is a common challenge, with behavioral responses ranging from hyperactivity to
disengagement. T1 noted:

\begin{quote}
``It's common to see a child suddenly lose interest and drift away mentally, forcing us to change the
session's direction entirely.''
\end{quote}

T2 emphasized the importance of seizing spontaneous moments:

\begin{quote}
``Sometimes, a patient unexpectedly says a word, and you have to seize that moment to build an
activity around it.''
\end{quote}

In physical therapy, the challenges revolve around adjusting exercises to match the patient's
fluctuating physical and cognitive states. T3 explained:

\begin{quote}
``Even with the same patient, what worked one day might not work the next because of fatigue or
changing motivation levels.''
\end{quote}

Maintaining patient motivation is critical, as T3 pointed out:

\begin{quote}
``Patients can quickly become discouraged if they feel they're not progressing, so it's essential to
keep them motivated.''
\end{quote}

These challenges highlight the need for adaptive tools that allow therapists to respond in real time
to patient needs.

\subsection{Adaptation Strategies}

To enhance therapy effectiveness, therapists use various adaptation strategies to maintain patient
engagement and motivation. T2 stressed the value of engaging tools:

\begin{quote}
``Educational games and visual aids help keep the session lively and prevent the patient from losing
focus.''
\end{quote}

T1 often uses toys and real-time session adjustments to tailor activities based on the child's
reactions. T3 added:

\begin{quote}
``If a patient is struggling, you sometimes need to simplify the activity. If they're doing well, you
might increase the complexity to keep them challenged.''
\end{quote}

Rhythm plays a crucial role in therapy, particularly for neurological conditions. T4 discussed using
tools like the metronome:

\begin{quote}
``We've been using the metronome for a long time, but a constant beat might not always suit real-life
scenarios.''
\end{quote}

For patients struggling with rhythm, devices emitting rhythmic vibrations are useful, as T4 explained:

\begin{quote}
``Devices for Parkinson's patients emit vibrations to help maintain their walking rhythm.''
\end{quote}

Positive reinforcement is another key strategy, as T2 noted:

\begin{quote}
``A simple word of encouragement, like `well done,' can make a big difference in keeping a patient
motivated.''
\end{quote}

Therapists also adjust activities to match patient progress. T3 mentioned using external rhythms in
Parkinson's therapy:

\begin{quote}
``Patients lose their temporal references and can no longer maintain a rhythm, so we provide an
external rhythm that their brain can no longer sustain.''
\end{quote}

Biofeedback tools are also effective, as T4 stated:

\begin{quote}
``Seeing their progress in real time on a screen motivates patients.''
\end{quote}

Tailored approaches are essential, especially in conditions like Parkinson's, where specific tools
help maintain functionality. T4 explained:

\begin{quote}
``Devices can help Parkinson's patients maintain legible handwriting as it tends to become
progressively smaller.''
\end{quote}

These strategies underscore the importance of flexible, adaptive practices in rehabilitation.

\subsection{Interaction Between Motor Gestures and Speech}

Motor gestures play a significant role in enhancing speech and communication in therapy. T3 observed:

\begin{quote}
``We all talk with our hands. It's spontaneous and helps with communication, even when words are hard
to come by.''
\end{quote}

In speech therapy, gestures reinforce verbal messages and aid memorization. T1 provided an example:

\begin{quote}
``I sometimes use objects like small pieces on a table to represent syllables. The child taps each
piece as they say the syllable, which helps them see and feel the rhythm.''
\end{quote}

T1 also noted the effectiveness of combining rhythm and gestures:

\begin{quote}
``Tapping out the rhythm while reciting something like the days of the week helps children remember
better because they associate the rhythm with the words.''
\end{quote}

In physical therapy, T3 mentioned using gestures alongside verbal instructions to help patients
synchronize actions with therapy rhythms. The integration of speech, gesture, and rhythm is
particularly effective for patients with both motor and verbal coordination challenges, enhancing
therapeutic outcomes.

\section{Interface Adaptations Based on Therapist Feedback}

Therapist feedback was vital in guiding adaptations to the rhythmic interface, ensuring it meets the
diverse needs of patients in both speech and physical therapy. Flexibility and personalization were
key priorities.

In response to the need for greater flexibility in therapy sessions, a dedicated web interface was
developed, requiring therapists to upload both an audio file (e.g., a word, phrase, or specific
exercise, recorded by the therapist or from existing material) and the corresponding text file via a
drag-and-drop process. Once the files are uploaded, the interface automatically processes them using
the same tools as described in Figure~\ref{fig:pipeline}, such as Vocal Remover and the Montr\'eal
Forced Aligner, to perform segmentation and synchronization of the audio with minimal user input.

In the final stage, the therapist can adjust the automatically generated timestamps by interacting
with a visual waveform representation of the audio. The black bars represent calculated timestamps,
which can be moved as needed to refine the exercise. Alternatively, the therapist may choose to leave
the timestamps unchanged if they are satisfied with the initial output.

The interface includes four main controls: a play button to listen to the current segment (starting
with the first segment between timestamps 1 and 2), a next button to skip to the following segment, a
third button to play the audio from the current timestamp to the end, and a download button that
generates a \texttt{.txt} file. This text file, organized with all the timestamps in the correct
format, is ready to be uploaded for patient use in subsequent sessions.

\begin{figure}[htbp]
  \centering
  \includegraphics[width=0.95\textwidth]{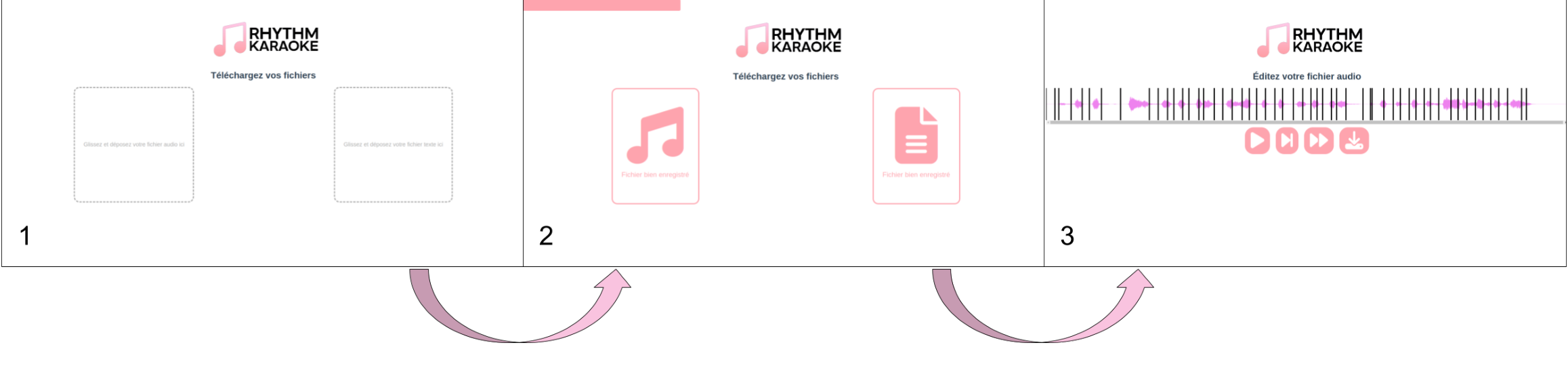}
  \caption{Workflow of the web interface for uploading and processing audio files. In (1), the
  drag-and-drop functionality is used to upload audio and text files. In (2), the files are
  automatically processed. In (3), the interface allows therapists to view and adjust the timestamps
  (represented as black bars) on the waveform. The control buttons below the waveform enable
  therapists to navigate through the segments or download the final timestamp file.}
  \label{fig:webui}
\end{figure}

The interface now includes a variety of exercises, such as repeating common words or numerical
sequences, and integrates words related to daily activities and school terms. Exercises are
categorized by speed and length, allowing therapists to adjust difficulty levels, making the tool
adaptable to varied patient needs.

User experience improvements were also made, enhancing the interface's intuitiveness and visual
clarity. These changes ensure the interface is both engaging for young patients and easy for
therapists to integrate into their practices. Overall, the refined rhythmic interface offers a
flexible, personalized solution that aligns with current therapeutic needs.

\begin{figure}[htbp]
  \centering
  \includegraphics[width=\textwidth]{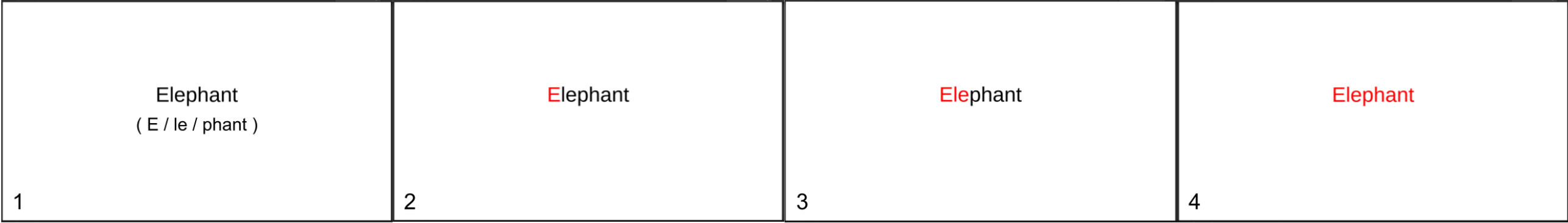}
  \caption{Illustration of the syllable decomposition and user interactions with the interface during
  the ``Elephant'' word repetition exercise. From left to right, images numbered 1 to 4 depict the
  progressive stages of the exercise. In (1), the word ``Elephant'' is displayed in black, with no
  syllables yet pronounced. In (2), after the user presses the button for the first syllable ``E,''
  it is highlighted in red, while the remaining syllables stay black. Simultaneously, the synthesized
  voice pronounces the syllable ``E,'' providing auditory feedback. In (3), the progression continues
  as the second syllable ``le'' is pressed, turns red, and is pronounced by the synthesized voice.
  Finally, in (4), the user completes the exercise by pressing the button for the third syllable
  ``phant,'' turning the entire word red and triggering the pronunciation of the final syllable. This
  visual and auditory sequence guides the user through each step, reinforcing coordination between
  gestures and the pronunciation of syllables, as well as ensuring real-time feedback with each
  interaction.}
  \label{fig:elephant}
\end{figure}

\section{Experience and Evaluation with Children}

Following interface modifications, therapy sessions were conducted with eight children
(Table~\ref{tab:children}) and T1.

\subsection{Motivation and Engagement}

During speech therapy sessions, the updated interface quickly captured the attention of P1 to P8,
sparking their curiosity. Even the most easily distracted children focused on the screen, exploring
its features with enthusiasm and hesitation. The exercises included simple words like ``spoon,''
``apple,'' and ``elephant'' (see Figure~\ref{fig:elephant}), as well as word series, poetry excerpts,
and phrases from the film \emph{The Jungle Book}.

As words appeared on the screen, the children were fascinated by the accompanying colors and sounds.
Some hesitated, gently tapping the screen, while others eagerly pressed the words. T1 observed:
``They're curious, but not always patient. Some of them get bored quickly if the word doesn't
entertain them enough.'' This variety in reactions allowed the exercises to be adjusted according to
each child's interests.

To better meet the children's needs, additional words were integrated into the interface based on
their preferences. For instance, after noticing that P2 responded especially well to animal names,
words like ``lion,'' ``elephant,'' and ``mouse'' were added. T1 explained: ``Being able to add words
based on their reactions was useful. It allowed us to personalize the sessions even more.''

However, while the children responded positively to the system, setting it up required preparation
time. Although T1 valued the flexibility of adding new words, this task could be challenging during
short sessions. ``It's a great tool, but it takes time to set up, and in a 30-minute session, every
minute counts,'' T1 noted.

\subsection{Evaluation of Progress}

T1 observed mixed results regarding the children's progress. While the interface generally improved
concentration and participation, some children struggled to maintain interest over time. ``Some kids
tune out pretty quickly if the exercise doesn't grab them right away,'' T1 remarked.

Adding new words helped to re-engage children who were losing focus. For example, when P1 seemed
impatient, offering a newly added word could reignite their interest. T1 added: ``Adding words based
on their interests helped us capture their attention better, even though sometimes it meant juggling
with time.''

When P3 successfully pronounced a new word, it was particularly rewarding. For instance, P3, who
struggled with ``apple,'' eventually managed to say it correctly after several repetitions, boosting
their confidence. ``Seeing them succeed gives them a boost,'' T1 commented.

\subsection{The Concept of Dose-Effect}

T1 emphasized the importance of pacing exercises to avoid exhausting the children's attention. ``Too
many exercises in a row, and they check out,'' T1 observed. The interface allowed for structuring
sessions into short, repeated sequences while staying attentive to signs of fatigue or disinterest.

Short but regular exercises, combined with the gradual introduction of new words, were favored to help
the children practice without feeling overwhelmed. T1 noted: ``Introducing a new word mid-session
often re-engages their attention.''

However, managing this flexibility in short sessions was challenging. T1 needed to balance the time
spent configuring the interface with direct interaction. ``It's great to personalize, but you have to
find a balance so it doesn't take up too much session time,'' T1 mentioned.

\subsection{Impact on Verbal Skills}

The interface positively impacted the children's verbal and motor skills, though results varied. Some
children repeated words with precision, while others rushed, sometimes without correct pronunciation.
``Some repeat just to finish faster,'' T1 observed, stressing the importance of encouraging quality
over quantity in repetitions.

In some cases, adding new words provided additional challenges, helping children progress further.
For example, P8, who struggled with certain sounds, showed significant improvement after more complex
words were introduced, allowing for targeted work on specific phonemes.

In conclusion, the experience with P1 to P8 demonstrated that the rhythmic interface is a promising
tool for improving motivation and language skills. While adjustments are needed to match each child's
pace and reactions, and while adding new words requires time, the observed progress in participation
and linguistic performance highlights the potential of this technology to enrich therapeutic
practices. A flexible approach tailored to individual needs is key to maximizing the tool's
effectiveness while managing session time efficiently.

\section{Discussion}

Our study confirmed some anticipated outcomes while revealing unexpected insights, particularly
regarding the cross-domain potential of integrating rhythm, gestures, and speech in therapy. As
expected, the rhythmic interface enhanced children's motivation and engagement, consistent with
existing research emphasizing rhythm's role in sustaining attention and facilitating learning. The
ability to customize exercises by adding new words based on individual preferences significantly
improved the flexibility and efficacy of the sessions, allowing therapists to adapt more precisely to
each child's needs.

One of the key findings was the successful adaptation of a motor rehabilitation tool to speech
therapy, taking advantage of the cognitive link between gestures and speech. This discovery aligns
with research by McNeill~\cite{mcneill2005gesture}, which demonstrated how gestures are integral to
language processing. Our findings extend this understanding, showing that a system initially designed
to enhance motor skills can also be applied to speech rehabilitation, leveraging the link between
gesture and verbal communication.

An additional insight from our results was the influence of rhythm on speech fluency. The controlled
tempo provided by the interface facilitated smoother speech production, consistent with findings on
slow speech rate techniques used in stuttering
therapy~\cite{ingham2001evaluation,lasalle2015slow}. This technique, by reducing cognitive load,
likely allowed children more time to articulate their speech more clearly, leading to improved verbal
output.

The multisensory nature of the interface, which combines auditory, visual, and motor stimuli, further
amplified the children's engagement. Studies have demonstrated that such multisensory inputs enhance
both learning and memory retention by stimulating multiple brain regions
simultaneously~\cite{macedonia2012gestures}. This capability to engage different senses in a
synchronized manner likely played a key role in maintaining the children's focus and facilitating
their learning process.

The interactive nature of the system, enhanced by real-time feedback, provided a powerful
reinforcement mechanism. Research shows that immediate auditory and visual feedback can accelerate
learning by reinforcing the association between actions and
outcomes~\cite{schlaug2008singing,schlaug2010singing}. The interface's feedback mechanism likely
helped the children solidify the connection between gestures and verbal expressions, improving their
overall speech performance.

One notable result was the increased verbal output observed in children using the interface. The
combination of interactivity and personalization, allowing for tailored exercises, may have
encouraged greater participation. This aligns with Asher's work~\cite{asher1968total}, which shows
that incorporating physical movement into language learning stimulates verbal production. The
rhythmic and interactive nature of our system may have created an environment conducive to more
spontaneous speech.

In terms of long-term benefits, research suggests that rhythmic and motor-based interventions can
have lasting effects on neural connectivity, supporting sustained improvements in both motor and
verbal skills~\cite{schlaug2008singing,schlaug2010singing}. Investigating whether the progress
observed during therapy sessions translates into durable gains in language acquisition and fluency
would be a valuable direction for future research.

Given these findings, the potential to apply this system in reverse---using speech to enhance motor
skills---deserves exploration. For instance, verbal cues could be incorporated into motor
rehabilitation exercises to strengthen the cognitive link between speech and movement, potentially
improving outcomes in stroke recovery or neurodegenerative diseases.

Moreover, the use of rhythmic gestures could be expanded into language learning environments. The
coordination of speech and movement might facilitate vocabulary acquisition and reinforce cognitive
associations, as demonstrated in previous studies on the role of motor activities in language
learning~\cite{macedonia2012gestures}. Our findings suggest that gesture-based rhythmic interfaces
could be developed to assist learners by synchronizing vocabulary learning with physical actions,
amplifying the benefits observed in methods like the Total Physical Response
approach~\cite{asher1968total}.

In terms of rehabilitation, our study supports the idea that integrating rhythm and gestures is
beneficial across various conditions, including aphasia, where patients face both motor and verbal
coordination challenges. Prior research has shown that gestures can aid speech recovery in aphasia
patients~\cite{sekine2013relationship}. The rhythmic interface could be further adapted to make
therapy more interactive and engaging, where patients use gestures to trigger words, strengthening
the connection between movement and language. Additionally, this approach could be explored in
contexts of neurostimulation, where the interaction between rhythmic stimulation and speech
production remains under-researched~\cite{schlaug2008singing,schlaug2010singing}.

Finally, the connection between gestures and speech could open new avenues in neuroeducation,
particularly in managing conditions like attention-deficit hyperactivity disorder. Integrating
rhythmic gestures into cognitive exercises could help channel children's motor energy while enhancing
focus and learning, as supported by studies on the impact of physical activity on executive
functions~\cite{gapin2010relationship}.

\section{Conclusions}

Our research explores the adaptation of a rhythmic interface initially designed for manual dexterity
rehabilitation, which we successfully applied to speech therapy. This adaptation reveals the
cognitive link between gestures and speech, enhancing motivation and therapeutic engagement in
children with speech disorders. Our study contributes to understanding the challenges of speech
therapy with children and can guide designers interested in incorporating novel technologies in this
field.

The study opens up possibilities for cross-domain applications, including speech improvement, motor
skills rehabilitation in neurorehabilitation contexts like stroke recovery, and educational settings
where synchronized gestures and speech could aid language learning and cognitive development.

Preliminary results suggest that the rhythmic interface can strengthen the connection between
gestures and verbal communication in speech therapy. Further studies are needed to validate these
findings in broader, long-term therapeutic contexts.

\bibliographystyle{plain}

\begin{thebibliography}{99}
\small

\bibitem{vocalremover}
Vocal Remover. \url{https://vocalremover.org/}. Accessed 2023-01-18.

\bibitem{andric2012gesture}
Michael Andric and Steven L. Small. 2012. Gesture's neural language. \emph{Frontiers in Psychology}
3 (2012), 99.

\bibitem{asher1968total}
James Asher. 1968. The total physical response method for second language learning. \emph{San Jose:
San Jose State College} (1968).

\bibitem{ashoori2015effects}
Aidin Ashoori, David M. Eagleman, and Joseph Jankovic. 2015. Effects of auditory rhythm and music on
gait disturbances in Parkinson's disease. \emph{Frontiers in Neurology} 6 (2015), 234.

\bibitem{bassett2015learning}
Danielle S. Bassett, Muzhi Yang, Nicholas F. Wymbs, and Scott T. Grafton. 2015. Learning-induced
autonomy of sensorimotor systems. \emph{Nature Neuroscience} 18, 5 (2015), 744--751.

\bibitem{bengtsson2009listening}
Sara L. Bengtsson, Fredrik Ullen, H. Henrik Ehrsson, Toshihiro Hashimoto, Tomonori Kito, Eiichi
Naito, Hans Forssberg, and Norihiro Sadato. 2009. Listening to rhythms activates motor and premotor
cortices. \emph{Cortex} 45, 1 (2009), 62--71.

\bibitem{boersma2022praat}
Paul Boersma and David Weenink. 1992--2022. Praat: doing phonetics by computer [Computer program].
Version 6.1.08, retrieved 5 December 2019 from \url{http://www.praat.org}.

\bibitem{braun2006using}
Virginia Braun and Victoria Clarke. 2006. Using thematic analysis in psychology. \emph{Qualitative
Research in Psychology} 3, 2 (2006), 77--101.

\bibitem{braun2016using}
Virginia Braun, Victoria Clarke, and Paul Weate. 2016. Using thematic analysis in sport and exercise
research. In \emph{Routledge Handbook of Qualitative Research in Sport and Exercise}. Routledge,
213--227.

\bibitem{braunjanzen2022rhythm}
Thenille Braun Janzen, Yuko Koshimori, Nicole M. Richard, and Michael H. Thaut. 2022. Rhythm and
music-based interventions in motor rehabilitation: current evidence and future perspectives.
\emph{Frontiers in Human Neuroscience} 15 (2022), 789467.

\bibitem{dayan2011neuroplasticity}
Eran Dayan and Leonardo G. Cohen. 2011. Neuroplasticity subserving motor skill learning.
\emph{Neuron} 72, 3 (2011), 443--454.

\bibitem{delrio2019hand}
Marta Sylvia Del Rio Guerra, Jorge Martin-Gutierrez, Renata Acevedo, and Sofia Salinas. 2019. Hand
gestures in virtual and augmented 3D environments for Down syndrome users. \emph{Applied Sciences}
9, 13 (2019), 2641.

\bibitem{dick2012gesture}
Anthony Steven Dick, Susan Goldin-Meadow, Ana Solodkin, and Steven L. Small. 2012. Gesture in the
developing brain. \emph{Developmental Science} 15, 2 (2012), 165--180.

\bibitem{fujii2014role}
Shinya Fujii and Catherine Y. Wan. 2014. The role of rhythm in speech and language rehabilitation:
The SEP hypothesis. \emph{Frontiers in Human Neuroscience} 8 (2014), 777.

\bibitem{gapin2010relationship}
Jennifer Gapin and Jennifer L. Etnier. 2010. The relationship between physical activity and executive
function performance in children with attention-deficit hyperactivity disorder. \emph{Journal of
Sport and Exercise Psychology} 32, 6 (2010), 753--763.

\bibitem{ghai2018effects}
Shashank Ghai and Ishan Ghai. 2018. Effects of rhythmic auditory cueing in gait rehabilitation for
multiple sclerosis: a mini systematic review and meta-analysis. \emph{Frontiers in Neurology} 9
(2018), 386.

\bibitem{gonzalez2021effects}
Samira Gonzalez-Hoelling, Carme Bertran-Noguer, Gloria Reig-Garcia, and Rosa Su\~ner-Soler. 2021.
Effects of a music-based rhythmic auditory stimulation on gait and balance in subacute stroke.
\emph{International Journal of Environmental Research and Public Health} 18, 4 (2021), 2032.

\bibitem{grafton1996localization}
Scott T. Grafton, Michael A. Arbib, Luciano Fadiga, and Giacomo Rizzolatti. 1996. Localization of
grasp representations in humans by positron emission tomography: 2. Observation compared with
imagination. \emph{Experimental Brain Research} 112 (1996), 103--111.

\bibitem{hikosaka2002central}
Okihide Hikosaka, Kae Nakamura, Katsuyuki Sakai, and Hiroyuki Nakahara. 2002. Central mechanisms of
motor skill learning. \emph{Current Opinion in Neurobiology} 12, 2 (2002), 217--222.

\bibitem{hostetter2019gesture}
Autumn B. Hostetter and Martha W. Alibali. 2019. Gesture as simulated action: Revisiting the
framework. \emph{Psychonomic Bulletin \& Review} 26 (2019), 721--752.

\bibitem{ingham2001evaluation}
Roger J. Ingham, Martin Kilgo, Janis C. Ingham, Richard Moglia, Heather Belknap, and Tracy Sanchez.
2001. Evaluation of a stuttering treatment based on reduction of short phonation intervals.
\emph{Journal of Speech, Language, and Hearing Research} (2001).

\bibitem{johansenberg2002attention}
Heidi Johansen-Berg and P. Matthews. 2002. Attention to movement modulates activity in sensori-motor
areas, including primary motor cortex. \emph{Experimental Brain Research} 142 (2002), 13--24.

\bibitem{keaybright2011cocreating}
Wendy Keay-Bright and J. G. Lewis. 2011. Co-creating tools for touch: Applying an
Inspire-Create-Play-Appropriate methodology for the ideation of therapeutic technologies. In
\emph{Proceedings of Include 2011}. Awarded Best Innovation in Inclusive Design.

\bibitem{kelly2015processing}
Spencer Kelly, Meghan Healey, Asli \"Ozy\"urek, and Judith Holler. 2015. The processing of speech,
gesture, and action during language comprehension. \emph{Psychonomic Bulletin \& Review} 22 (2015),
517--523.

\bibitem{kendon2004gesture}
Adam Kendon. 2004. \emph{Gesture: Visible Action as Utterance}. Cambridge University Press.

\bibitem{kita2017how}
Sotaro Kita, Martha W. Alibali, and Mingyuan Chu. 2017. How do gestures influence thinking and
speaking? The gesture-for-conceptualization hypothesis. \emph{Psychological Review} 124, 3 (2017),
245.

\bibitem{kotz2010cortical}
Sonja A. Kotz and Michael Schwartze. 2010. Cortical speech processing unplugged: a timely
subcortico-cortical framework. \emph{Trends in Cognitive Sciences} 14, 9 (2010), 392--399.

\bibitem{lasalle2015slow}
Lisa R. LaSalle. 2015. Slow speech rate effects on stuttering preschoolers with disordered phonology.
\emph{Clinical Linguistics \& Phonetics} 29, 5 (2015), 354--377.

\bibitem{luo2023studying}
Lu Luo and Lingxi Lu. 2023. Studying rhythm processing in speech through the lens of auditory-motor
synchronization. \emph{Frontiers in Neuroscience} 17 (2023), 1146298.

\bibitem{macedonia2012gestures}
Manuela Macedonia and Katharina von Kriegstein. 2012. Gestures enhance foreign language learning.
\emph{Biolinguistics} 6, 3--4 (2012), 393--416. \url{https://doi.org/10.5964/bioling.8931}

\bibitem{mainka2014rhythmic}
Stefan Mainka and Grit Mallien. 2014. Rhythmic speech cueing (RSC). \emph{Handbook of Neurologic
Music Therapy} (2014), 150--160.

\bibitem{mcauliffe2017montreal}
Michael McAuliffe, Michaela Socolof, Sarah Mihuc, Michael Wagner, and Morgan Sonderegger. 2017.
Montreal Forced Aligner: Trainable text-speech alignment using Kaldi. In \emph{Interspeech 2017}.
498--502.

\bibitem{mcneill2005gesture}
David McNeill. 2005. \emph{Gesture and Thought}. University of Chicago Press.
\url{https://doi.org/10.7208/chicago/9780226514642.001.0001}

\bibitem{mcneill2011gestures}
David McNeill and Susan Duncan. 2011. Gestures and growth points in language disorders. In \emph{The
Handbook of Psycholinguistic and Cognitive Processes}. Psychology Press, 663--685.

\bibitem{mortenson2015development}
Ben W. Mortenson, Louise Demers, Marcus J. Fuhrer, Jeffrey W. Jutai, James Lenker, and Frank
DeRuyter. 2015. Development and preliminary evaluation of the caregiver assistive technology outcome
measure. \emph{Journal of Rehabilitation Medicine} 47, 5 (2015), 412--418.

\bibitem{nieuwboer2007cueing}
Alice Nieuwboer, Gert Kwakkel, Lynn Rochester, Diana Jones, Erwin van Wegen, Anne Marie Willems,
Fabienne Chavret, Victoria Hetherington, Katherine Baker, and Inge Lim. 2007. Cueing training in the
home improves gait-related mobility in Parkinson's disease: the RESCUE trial. \emph{Journal of
Neurology, Neurosurgery \& Psychiatry} 78, 2 (2007), 134--140.

\bibitem{nirme2024early}
Jens Nirme, Agneta Gulz, Magnus Haake, and Marianne Gullberg. 2024. Early or synchronized gestures
facilitate speech recall---a study based on motion capture data. \emph{Frontiers in Psychology} 15
(2024), 1345906.

\bibitem{nombela2013into}
Cristina Nombela, Laura E. Hughes, Adrian M. Owen, and Jessica A. Grahn. 2013. Into the groove: can
rhythm influence Parkinson's disease? \emph{Neuroscience \& Biobehavioral Reviews} 37, 10 (2013),
2564--2570.

\bibitem{patel2014evolutionary}
Aniruddh D. Patel. 2014. The evolutionary biology of musical rhythm: was Darwin wrong? \emph{PLoS
Biology} 12, 3 (2014), e1001821.

\bibitem{peelle2012neural}
Jonathan E. Peelle and Matthew H. Davis. 2012. Neural oscillations carry speech rhythm through to
comprehension. \emph{Frontiers in Psychology} 3 (2012), 320.

\bibitem{pulvermuller2005brain}
Friedemann Pulverm\"uller. 2005. Brain mechanisms linking language and action. \emph{Nature Reviews
Neuroscience} 6, 7 (2005), 576--582.

\bibitem{robinson2019game}
Ciar\'an Robinson. 2019. \emph{Game Audio with FMOD and Unity}. Routledge.

\bibitem{rosenkranz2007motorcortical}
Karin Rosenkranz, Aaron Williamon, and John C. Rothwell. 2007. Motorcortical excitability and
synaptic plasticity is enhanced in professional musicians. \emph{Journal of Neuroscience} 27, 19
(2007), 5200--5206.

\bibitem{ruud2023patient}
Torleif Ruud, Ingrid Kyte Fjellestad, and Ketil Hanssen-Bauer. 2023. Patient experiences in
psychiatric departments for the elderly (PEPDE): development, properties, and use of a brief
questionnaire. \emph{BMC Psychiatry} 23, 1 (2023), 173.

\bibitem{schaefer2014auditory}
Rebecca S. Schaefer. 2014. Auditory rhythmic cueing in movement rehabilitation: findings and possible
mechanisms. \emph{Philosophical Transactions of the Royal Society B: Biological Sciences} 369, 1658
(2014), 20130402.

\bibitem{schlaug2008singing}
Gottfried Schlaug, Sarah Marchina, and Andrea Norton. 2008. From singing to speaking: Why singing may
lead to recovery of expressive language function in patients with Broca's aphasia. \emph{Music
Perception} 25, 4 (2008), 315--323.

\bibitem{schlaug2010singing}
Gottfried Schlaug, Andrea Norton, Sarah Marchina, Lauryn Zipse, and Catherine Y. Wan. 2010. From
singing to speaking: facilitating recovery from nonfluent aphasia. \emph{Future Neurology} 5, 5
(2010), 657--665.

\bibitem{sekine2013relationship}
Kazuki Sekine and Miranda L. Rose. 2013. The relationship of aphasia type and gesture production in
people with aphasia. \emph{American Journal of Speech-Language Pathology} (2013).

\bibitem{skipper2007speech}
Jeremy I. Skipper, Susan Goldin-Meadow, Howard C. Nusbaum, and Steven L. Small. 2007.
Speech-associated gestures, Broca's area, and the human mirror system. \emph{Brain and Language} 101,
3 (2007), 260--277.

\bibitem{teremetz2015novel}
Maxime T\'er\'emetz, Florence Colle, Sonia Hamdoun, Marc A. Maier, and P\r{a}vel G. Lindberg. 2015. A
novel method for the quantification of key components of manual dexterity after stroke.
\emph{Journal of NeuroEngineering and Rehabilitation} 12 (2015), 1--16.

\bibitem{teremetz2023efficacy}
Maxime T\'er\'emetz, Sonia Hamdoun, Florence Colle, Elo\"ise Gerardin, Claire Desvilles, Loic
Carment, Sylvain Charron, Macarena Cuenca, David Calvet, Jean-Claude Baron, et al. 2023. Efficacy of
interactive manual dexterity training after stroke: a pilot single-blinded randomized controlled
trial. \emph{Journal of NeuroEngineering and Rehabilitation} 20, 1 (2023), 93.

\bibitem{thaut2010rhythmic}
Michael H. Thaut and Mutsumi Abiru. 2010. Rhythmic auditory stimulation in rehabilitation of movement
disorders: a review of current research. \emph{Music Perception} 27, 4 (2010), 263--269.

\bibitem{thaut1996rhythmic}
Michael H. Thaut, Gerald C. McIntosh, Ruth R. Rice, Robert A. Miller, Julie Rathbun, and John M.
Brault. 1996. Rhythmic auditory stimulation in gait training for Parkinson's disease patients.
\emph{Movement Disorders} 11, 2 (1996), 193--200.

\bibitem{zhao2023tms}
Wanying Zhao. 2023. TMS reveals a two-stage priming circuit of gesture-speech integration.
\emph{Frontiers in Psychology} 14 (2023), 1156087.

\bibitem{zumbansen2014melodic}
Anna Zumbansen, Isabelle Peretz, and Sylvie H\'ebert. 2014. Melodic intonation therapy: back to
basics for future research. \emph{Frontiers in Neurology} 5 (2014), 7.

\end{thebibliography}

\appendix

\section{Interview Protocol}
\label{app:protocol}

\subsection*{Introduction to the Interview}

I am currently developing a rhythm-based system that allows users to control the rhythm of
synthesized vocal phrases through finger tapping. The system is being designed for potential use in
therapeutic settings, with the aim of supporting both motor rehabilitation and speech therapy by
incorporating rhythmic exercises. The purpose of this interview is to gather your professional
insights on how such a system could be integrated into existing therapeutic practices and to
understand your perspective on the potential benefits or challenges it may present. Additionally, I
am interested in learning more about the current therapeutic methods you use, to better inform the
system's development and ensure it aligns with the practical needs of clinicians and patients.

\subsection*{Part 1: Questions about Current Practice}

\begin{enumerate}[leftmargin=2em]
  \item How are speech, rhythm, and gestures currently integrated into your practice?
  \item What types of content do you typically use during your sessions?
  \item What challenges do you encounter in your sessions related to speech, rhythm, and gestures?
\end{enumerate}

\subsection*{Part 2: Demonstration}

The demonstration of the interface takes place in this part of the interview.

\subsection*{Part 3: Questions about the Interface}

\emph{A: Interface Feedback and Current Use.}

\begin{enumerate}[leftmargin=2em]
  \item What are your thoughts on the interface? What do you see as its advantages and disadvantages?
  \item Can you provide examples of specific situations where you think this system could be
        particularly beneficial for patients?
  \item Can you think of other types of content that could be integrated into this system to improve
        it?
\end{enumerate}

\emph{B: Evaluation and Testing with Patients.}

\begin{enumerate}[leftmargin=2em]
  \item Would you be interested in testing this system with your clients/patients to evaluate its
        effectiveness and impact in a real-world context? If so, what steps and criteria would you
        suggest for conducting such a test?
  \item What success indicators or metrics would you consider most relevant for evaluating the
        effectiveness of this system?
  \item Were the questions I asked relevant? Do you feel that anything was missing? Do you have any
        additional comments or suggestions?
\end{enumerate}

\clearpage
\newgeometry{top=2cm,bottom=2cm,left=2.5cm,right=2.5cm}
\section{Challenges and Adaptive Strategies in Rehabilitation Sessions}
\label{app:challenges}

{\scriptsize
\setlength{\tabcolsep}{4pt}
\setlength{\LTpre}{4pt}
\setlength{\LTpost}{0pt}
\renewcommand{\arraystretch}{1.02}
\begin{longtable}{P{1.8cm}P{3.1cm}P{3.1cm}P{3.1cm}P{3.4cm}}
  \caption{Challenges, successful strategies, and adaptations observed during speech and physical
  therapy sessions.}
  \label{tab:challenges}\\
    \toprule
    \textbf{Theme} & \textbf{Specific Challenge} & \textbf{What Worked Well} &
    \textbf{What Did Not Work} & \textbf{Adaptation Strategy} \\
    \midrule
  \endfirsthead
    \toprule
    \textbf{Theme} & \textbf{Specific Challenge} & \textbf{What Worked Well} &
    \textbf{What Did Not Work} & \textbf{Adaptation Strategy} \\
    \midrule
  \endhead
    \bottomrule
  \endfoot

    Speech Therapy
      & Maintaining child engagement despite distractions
      & Seizing spontaneous moments of speech (e.g., when the child unexpectedly speaks)
      & Sudden loss of focus, hyperactivity, disengagement mid-session
      & Real-time adjustments using toys and educational games to maintain engagement. Use of rhythm
        and visual aids for consistency. \\
    \addlinespace[2pt]
      & Handling wide range of behavioral responses (hyperactivity to disengagement)
      & Adjusting activities based on child's reaction, making session dynamic and flexible
      & Some methods losing effectiveness mid-session if child loses focus or gets bored
      & Dynamic shift in session, introducing new stimuli like toys or objects representing syllables
        to re-engage the child. \\
    \addlinespace[2pt]
      & Using rhythm and gestures to reinforce speech
      & Combining rhythm with gestures (e.g., tapping objects to syllables) to improve memory
      & Repetitive rhythm may become less effective if overused in long sessions
      & Incorporating both auditory (rhythm) and physical (gestures) cues to reinforce learning,
        allowing breaks to re-introduce focus. \\
    \midrule

    Physical Therapy
      & Adjusting exercises to fluctuating physical and cognitive states (fatigue, motivation)
      & Simplifying or complicating exercises in real time based on patient performance
      & Exercises too difficult when patients are tired, leading to discouragement
      & Constant adaptation by simplifying when necessary, using positive feedback to maintain
        motivation. \\
    \addlinespace[2pt]
      & Keeping patients motivated throughout sessions
      & Positive reinforcement (e.g., verbal encouragement) and use of external rhythmic cues (e.g.,
        metronome, devices for Parkinson's)
      & Patients becoming demotivated if they don't perceive progress
      & Providing biofeedback tools showing real-time progress, helping patients visually understand
        their improvements. \\
    \addlinespace[2pt]
      & Balancing cognitive load with physical tasks
      & Breaking down complex exercises into smaller, manageable parts
      & Overloading patients with tasks that are too complex or lengthy
      & Simplifying instructions, providing extra time for task transitions, reinforcing small
        victories. \\
    \midrule

    Motor-Speech Integration
      & Using motor gestures to aid verbal expression
      & Gestures (e.g., hand movements) help reinforce and enhance verbal memory (linking movement to
        rhythm)
      & Difficulty synchronizing speech and motor gestures in some patients with advanced conditions
      & Providing physical objects to represent concepts (e.g., syllables) and using rhythm to
        support coordination. \\
    \addlinespace[2pt]
      & Synchronizing verbal instructions with physical tasks
      & Combining verbal cues with synchronized motor actions improves speech recall
      & Unsynchronized gestures can lead to confusion or failure to recall verbal cues
      & Breaking down tasks, offering more time for synchronization, and giving real-time feedback
        during sessions. \\
    \midrule

    General Therapy Strategies
      & Adjusting methods to different patient needs during the session
      & Real-time adaptability of tools, switching between methods depending on patient response
      & Rigid tools without adaptive features can impede flexibility
      & Using adaptive tools like metronomes with variable rhythms, biofeedback for real-time
        progress tracking. \\
    \addlinespace[2pt]
      & Incorporating rhythm in therapy for neurological conditions (e.g., Parkinson's)
      & Devices emitting rhythmic vibrations to help patients maintain walking or other activities
      & Monotonous rhythms may fail to engage patients long-term
      & Using devices that emit varied rhythmic cues, offering flexibility depending on patient's
        needs (e.g., adjusting speed or intensity). \\
    \addlinespace[2pt]
      & Adapting educational tools in therapy
      & Interactive, engaging tools (e.g., games, toys) help maintain focus and motivation
      & Standard tools become ineffective if they don't adapt to patient progression or mood changes
      & Use of customizable, interactive tools that adapt to the patient's real-time reactions and
        needs during therapy. \\
\end{longtable}
}
\restoregeometry

\end{document}